\documentclass[preprint]{aastex62}

\accepted{11 December 2020}
\submitjournal{AJ}

\shorttitle{VY CMa Mass Loss History}
\shortauthors{Humphreys et al.}

\begin{document}

\title{The Mass-Loss History of the Red Hypergiant VY CMa\footnote{Based on observations made with the NASA/ESA Hubble Space Telescope which is operated by the Association of Universities for Research in Astronomy, Inc., under NASA contract NAS 5-26555.}}

\correspondingauthor{Roberta Humphreys}
\email{roberta@umn.edu}

\author{Roberta M. Humphreys}
\affiliation{Minnesota Institute for Astrophysics,  
University of Minnesota,
Minneapolis, MN 55455, USA}
\nocollaboration

\author{Kris Davidson}
\affiliation{Minnesota Institute for Astrophysics,
University of Minnesota,
Minneapolis, MN 55455, USA}
\nocollaboration

\author{A. M. S. Richards}
\affiliation{Jodrell Bank,
Department of Physics and Astronomy, University of Manchester, UK}  
\nocollaboration

\author{L. M. Ziurys}
\affiliation{Departments of Astronomy and Chemistry,
University of Arizona}
\nocollaboration

\author{Terry J. Jones}
\affiliation{Minnesota Institute for Astrophysics,
University of Minnesota,
Minneapolis, MN 55455, USA}
\nocollaboration


\author{Kazunori Ishibashi}
\affiliation{Graduate School of Science, Nagoya University, Nagoya, 464-8602, Japan}
\nocollaboration



\begin{abstract}
	Imaging and spectroscopy of the knots, clumps, and extended arcs in the complex ejecta of VY CMa confirm a record of high mass loss events over the past few hundred years. {\it HST/STIS} spectroscopy of numerous small knots close to the star allow us to measure their radial velocities from the strong K I emission and determine their  separate motions, spatial orientations, and time since ejecta. Their ages concentrate around 70, 120, 200 and 250 years ago. A K I emission knot only 50 mas from the star ejected as recently as 1985 -- 1995 may  
coincide with an H$_{2}$O maser.  {\it Comparison with VY CMa's  historic light curve from 1800 to the present, shows several knots with ejection times that correspond with extended periods of variability and deep minima.  The similarity of this correspondence in VY CMa with the remarkable recent dimming of Betelgeuse and an outflow of gas is  apparent.} The evidence for similar outflows from the surface of a  more typical red supergiant suggests that discrete ejections are more common and surface or convective activity is a major source of mass loss for red supergiants.  
\end{abstract}


\section{Introduction} \label{sec:intro}

The famous red hypergiant VY CMa is one  of a small class of evolved 
massive stars that represent a short-lived evolutionary phase
characterized by extensive circumstellar ejecta, asymmetric ejections, and multiple high mass loss events. VY CMa is an extreme case even among the rare 
hypergiant stars. The three dimensional morphology of its  
complex circumstellar ejecta \citep{RMH2007,TJJ} and the massive, dusty condensations revealed by
ALMA sub-millimeter observations \citep{OGorman,Vlemmings2017,Kaminski,Asaki}
confirm a record of numerous high mass loss
events over the past few hundred years driven by localized, large 
scale instabilities on its surface. VY CMa is one of the most important examples for
understanding episodic  mass loss and the role of 
possible surface activity and magnetic fields.

Ground-based optical spectroscopy of the central 
star and ejecta
has been limited to 0$\farcs$8  spatial resolution  \citep{RMH2005}. 
Several small dusty knots and filaments  
within one arc second of the star very likely represent the most recent mass loss events.
We therefore obtained {\it HST/STIS} observations to probe  VY CMa's innermost 
ejecta and numerous small, diffuse condensations to the south and southwest of the 
star. 

Our spectra of the  small knots and filaments closest to the central star 
yielded a surprising result reported in our earlier Letter  \citep{RMH19}, hereafter Paper I. Very strong K I emission line formed by resonant scattering \citep{RMH2005} and 
 the TiO and VO molecular spectra, long assumed to form in a dusty circumstellar shell, actually originate in a few small clumps, a few 100 AU from the star. 
 Based on their motions, we found that they were ejected about one hundred 
 years ago and very likely represent VY CMa's most recent active period.
The K I lines are 10 to 20 times stonger in these nearest ejecta than on the
star itelf.  The excitation of these strong  atomic and molecular features 
present in the small knots therefore present an astrophysical problem. We 
showed that the clumps must have a nearly clear line of sight to the star's 
radiation implying major gaps or holes in the circumstellar medium perhaps formed by large-scale activity.  

In this second paper,  we focus on the spectra of the knots and clumps  
 visible in the images to the south and southwest of the star, the S knots, SW knots  and the SW Clump \citep{RMH2005,RMH2007}.  In the next section we describe our {\it HST/STIS} observations followed by a brief discussion of the spectrum of
the central star and the  presence of a small emission knot very close to 
the star possibly associated with H$_{2}$O maser emission. In \S {4}, the numerous K I emission peaks along the slits, 
their identification with the diffuse knots, their ejection ages, and the 
geometry of the SW Clump are discussed. In the final section we summarize the 
recent mass loss history of VY CMa. We find that numerous ejections over the past 200 years correspond to extended periods of major dimming in VY CMa's light curve reminiscent of  Betelgeuse's recent unexpected behavior.

\section{{\it HST/STIS} Observations} \label{sec:obs}

The {\it HST/STIS} observations were planned in two visits to allow  two 
different 
slit orientations sampling  the small knots and filaments immediately to 
the west and east of the star with three slit positions (Paper I), and to 
separate the 
individual knots to the south and southwest of the star with three slit positions as shown 
in Figure 1. The central star was observed in each visit.  

We used the STIS/CCD with the G750M grating at tilts 7795 and 8561, with respective spectral resolutions of 6900 and 7600, to measure the K I emission doublet  
and the near-infrared Ca II absorption triplet  and other absorption lines in the 8500{\AA} region.
The K I lines, formed by resonant scattering, are the strongest emission 
features in VY CMa and are the best tracers of the gas. The strong Ca II lines
 however are reflected  by dust, and are both redshifted and broadened by the dust scattering. The 52x0.1 slit was used for the small knots and filaments to the west and east of the star.    
The 0$\farcs$2 slit width was chosen for the knots south of the star, discussed here, to enclose their somewhat more diffuse, extended  
structures and to account for their motions measured previously \citep{RMH2007}.
These {\it HST/STIS} observations were 
completed on 2018 January 5 and February 11.

\begin{figure}[h]  
\figurenum{1}
\epsscale{0.8}
\plotone{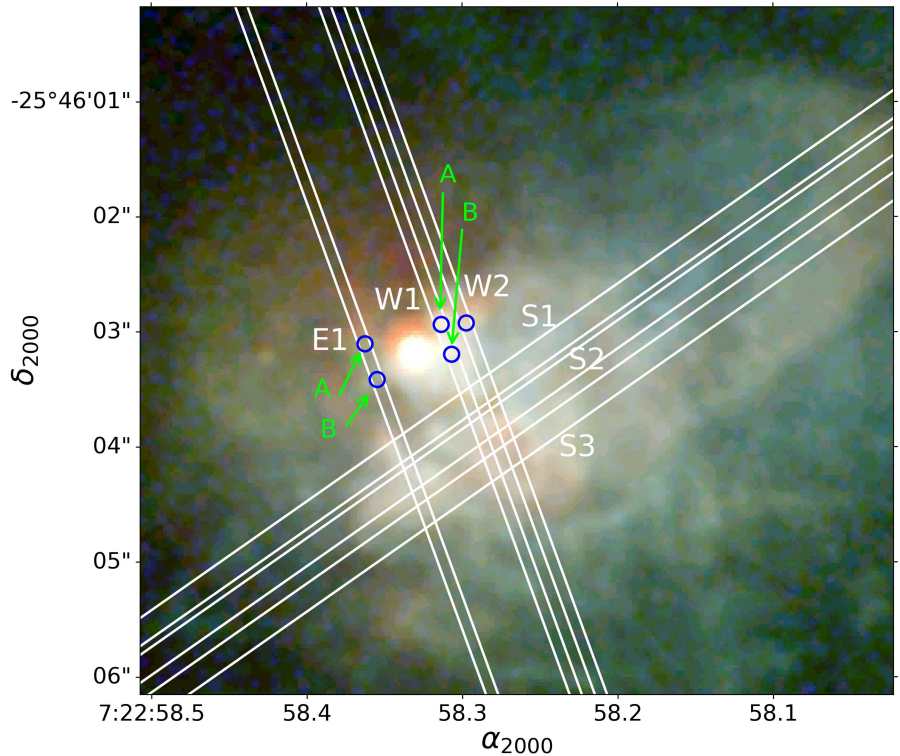}
\caption{Composite image of VY CMa from Paper I showing positions of the slits  and knots close to the star discussed in that paper.}
\label{fig:slits}   
\end{figure}

At $\lambda$ $\approx$ 7000 {\AA}, the basic spectral resolution of STIS is slightly better than 0$\farcs$1 (FWHM of the PSF). Since this is only about 2 CCD pixels we use the sub-pixel modeling technique developed 
for the eta Car Treasury programs \citep{Davidson} to assure a consistent data PSF and to take full advantage of the HST's 
spatial resolution along the slit. Consequently, the  ''sampling''scale in the processed 
spectra is  0$\farcs$0253 per pixel, rather than the CCD's 0$\farcs$05 scale.  Contemporary flat-field images were also 
obtained  
to correct for fringing in the red and  were  processed in the same way 
as the science images and  normalized to be used as contemporary fringe 
flat-field images.

The different velocities measured in the spectrum of VY CMa are discussed in \citet{RMH2005}. It is well established that the absorption and emission line velocities deviate from the systemic velocity inferred from the masers. As in previous papers, we adopt VY CMa's K I emission velocity as our reference frame. The velocity of the K I emission measured in the 0$\farcs$1 slit, 55 km s$^{-1}$ (Table 1) differs from that used in our 2005 and 2007 papers due to strong K I emission from compact knots in the wider slit in the groundbased spectra,  see  Paper I. In addition to the redshift and broadening due to scattering by dust, the absorption lines are also affected by the ``moving mirror'' effect.  

\section{The Central Star}  

With the two different slit widths as well as their separate orientations, we 
have two sample sizes of the central star's wind and innermost ejecta. The 0$\farcs$1 and 0$\farcs$2 slit widths correspond to $\approx$ 120 and 240 AU respectively at the 1.2 kpc distance of VY CMa \citep{Zhang}.  VY CMa's dust 
formation zone is expected at about 100 AU \citep{RMH2005}. The star's spectrum with the 0$\farcs$2 slit therefore samples more of the  dusty ejecta closest to the star.  Its spectrum of the K I region is shown in Figure 2, together 
with the  spectrum from
the same spectral region  obtained with the $0\farcs$1 slit. The same  width 
was used for the extractions.  Stronger K I emission and also weak molecular
emission is beginnng to appear in the spectrum with the wider slit width. For example, the two weak emission features are due to TiO, Paper I. The K I emission lines
 from the wider slit also show a small shift to somewhat longer wavelengths 
 due to the resonant scattering. The absorption lines, including the Ca II triplet plus lines of Fe I and Ti I, do not show a similar redshift relative to the narrower slit. Since these lines are seen via scattered light, the lack of any relative motion suggests very little expansion of the dusty inner  ejecta. The radial velocities measured on the star with the two slits are summarized in Table 1.

\begin{deluxetable*}{llll}
\tabletypesize{\footnotesize}
\tablenum{1}
\tablecaption{Heliocentric Velocities of the Star and Inner Knot}
\tablewidth{0pt}
\label{tab:Star}
\tablehead{
\colhead{Object} & 
\colhead{Avg K I em Vel} & 
\colhead{K I abs Vel} & 
\colhead{Avg Abs. Lines Vel.} \\ 
\colhead{slit} &
\colhead{km s$^{-1}$}  &
\colhead{km s$^{-1}$}  &
\colhead{km s$^{-1}$}   
}
\startdata 
Star 0$\farcs$1  &  55 $\pm$ 4   & 7.0   & 67.1 $\pm$ 11 (21) \\ 
Star 0$\farcs$2  &  68.7 $\pm$ 4 & 10.7  & 60.7 $\pm$ 11 (18) \\
Inner Knot 0$\farcs$1  & 47.6 $\pm$ 2 &  4.3 &  65.7 $\pm$ 8.4 (19)   
\enddata
\end{deluxetable*}

\begin{figure}
\figurenum{2}
\epsscale{1.3}
\plottwo{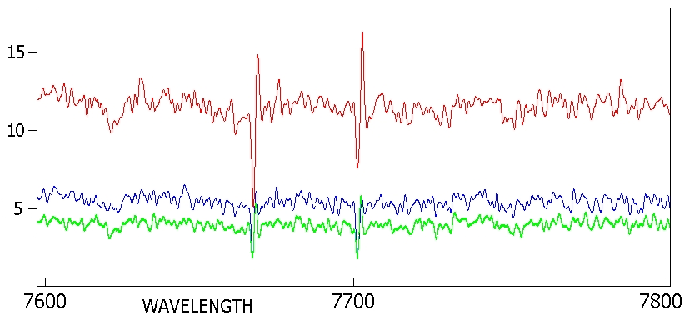}{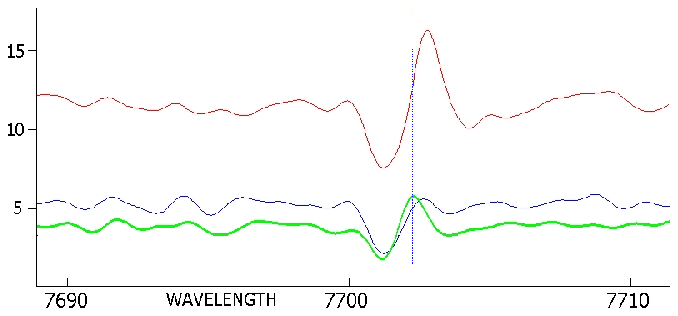}
\caption{Upper: Spectra of the K I lines on the star in the 0\farcs1 slit in blue and the 0\farcs2 slit in red. The K I  lines from the small emission knot in the narrow slit just to the southwest of the star are shown in green. Lower: A closeup of the $\lambda$ 7701 K I profile. The vertical line marks the velocity of the inner knot just to the southwest of the star. The flux calibrated spectra are in units of 10$^{-14}$ Watts cm$^{-2}$ {\AA}$^{-1}$ vs the vacuum wavelength in {\AA}} 
\label{fig:Star}   
\end{figure}


The narrow 0$\farcs$1 slit also revealed a K I emission peak separated only 2 pixels or $0\farcs$05 from
the center of the star's spectrum. Given the slit's orientation on the star with PA 20 degree, 
this feature would be located just to its south-southwest. It is very likely a small knot of nebulosity or ejecta, the closest to the star. The  K I spectrum of this "inner knot"  is also shown in Figure 2.  Its K I emission peaks have an average velocity  of 47.6 km s$^{-1}$, slightly blue-shifted by -7 km s$^{-1}$  relative to the star's K I velocity (Table 1), and similar to the velocities measured for the small knots to the west and east of the star in Paper I. It is too close to the star to be visible in any of the HST images \citep{RMH2007}.  We therefore do not have a measured proper motion for it and its tranverse velocity is not known. Its negative radial component implies that it is in front of the plane of the sky. Its distance corresponds to only 60 AU from the star, within the star's dusty envelope.  If we assume that its total velocity is similar to those for the knots just to the west of the star in Paper I (20 -- 30 km s$^{-1}$),  then this knot was ejected only $\approx$  10 to 15 years before our first epoch 1999 images i.e., in 1985--1990.   

There is an interesting possible correlation of this inner knot with H$_{2}$O 
maser emission peaks within 100 mas shown in  Figure 17 in \citet{Asaki}. Most are located to the south of the star. The figure shows a velocity data cube with LSRK radial velocities. Our velocities are Heliocentric. When corrected to the LSR, the K I emission  velocity of the inner knot is 31.1 km s$^{-1}$ \footnote{The correction to the LSR depends on the adopted Solar motion. We use the Solar motion from \citet{Binney}.  LSRK (Local Standard of Rest (Kinematic) used for the radio observations is based on the average velocity of stars near the Sun.}.  Two of the published frames show maser peaks within 50 mas to the south and southwest of the star with velocities of 29.8 and 37.8 km s$^{-1}$ which may be coincident with inner the knot.  For  a direct comparison, we show the distribution of the maser emission  at 30.5 km s$^{-1}$ with the  0$\farcs$1 slit and the position of the inner knot superposed in Figure 3. 

In an earlier study covering the entire envelope of VY CMa, \citet{Ziurys}  
concluded that the molecular maser and thermal emission and the K I emission traced the same outflows. We therefore suggest that the optical K I emission inner knot may coincide with the H$_{2}$O masers at similar velocities.  

\begin{figure}[h] 
\figurenum{3}
\epsscale{0.5} 
\plotone{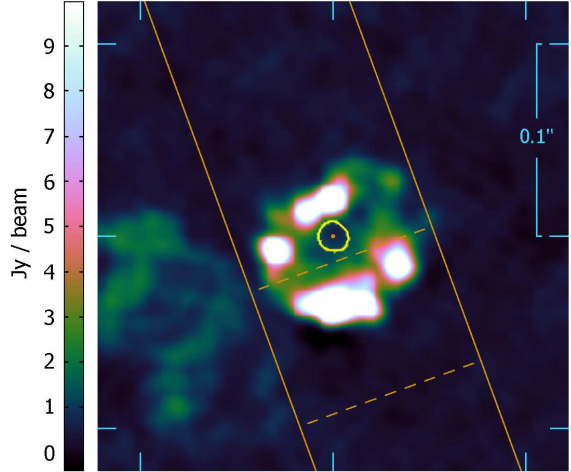}
 \caption{The H$_{2}$O 658 GHz maser emission peaks within 100mas of the star. The color scale is the 658 GHz maser in a 2.2 km s$^{-1}$ (averaged) channel at 30.5 km/s, with channel peak 34.72 Jy/beam, and  resolution (10x9) mas.  The yellow contour encloses 50\% of the continuum emission at 669 GHz; the dot  marks the 99\% contour of the 107 mJy/beam peak taken as the position of the star,
at  7h 22m 58.326s, -25deg 46' 3.04".  The parallel  orange lines map
    the position of the {\it HST/STIS} 0$\farcs$1  wide slit with PA 20 degrees.
The two dashed lines bracket the 3 row/pixel extraction width for the
spectrum of the K I emission feature that was centered at 0.05 arcsec
    from the star.  The 658 GHZ image was provided by A.M.S.Richards, see \citet{Asaki}.}
\end{figure}

\section{The Knots and Clumps South and Southwest of the Star}

\vspace{2mm}  

Figure 4 shows a close up of the positions of the three South(S) slits superposed on the F656N image. After acquisition, with the slit rotated to PA 125 degree, the star was centered on the slit using a peakup  in the red and an exposure taken.

The telescope was then offset to  the positions shown in Figure 4, -0\farcs716(S1), -1\farcs088(S2) and -1\farcs425(S3). The 2-D spectrum of the K I lines in  the three slits is shown in Figure 5  and the profiles corresponding to the K I emission peaks are shown in Figure 6.  The same 5 pixel wide extractions along the slit were used  for all of the spectra. The emission peaks were cross identified with the diffuse knots and condensations visible in the image by comparing their measured positions along the slit with the positions of the  emission peaks knowing the scale of the spectra and images. The identified knots are labeled in Figure 6 and listed in Table 2.  We follow the naming convention adopted
in \citet{RMH2007}, but note that some of the features listed here were not
included in that paper. They are identified as new measurements in the 
footnotes. In  a couple of cases there is no obvious identification with a knot.


\begin{figure}
\figurenum{4}
\epsscale{0.6}
\plotone{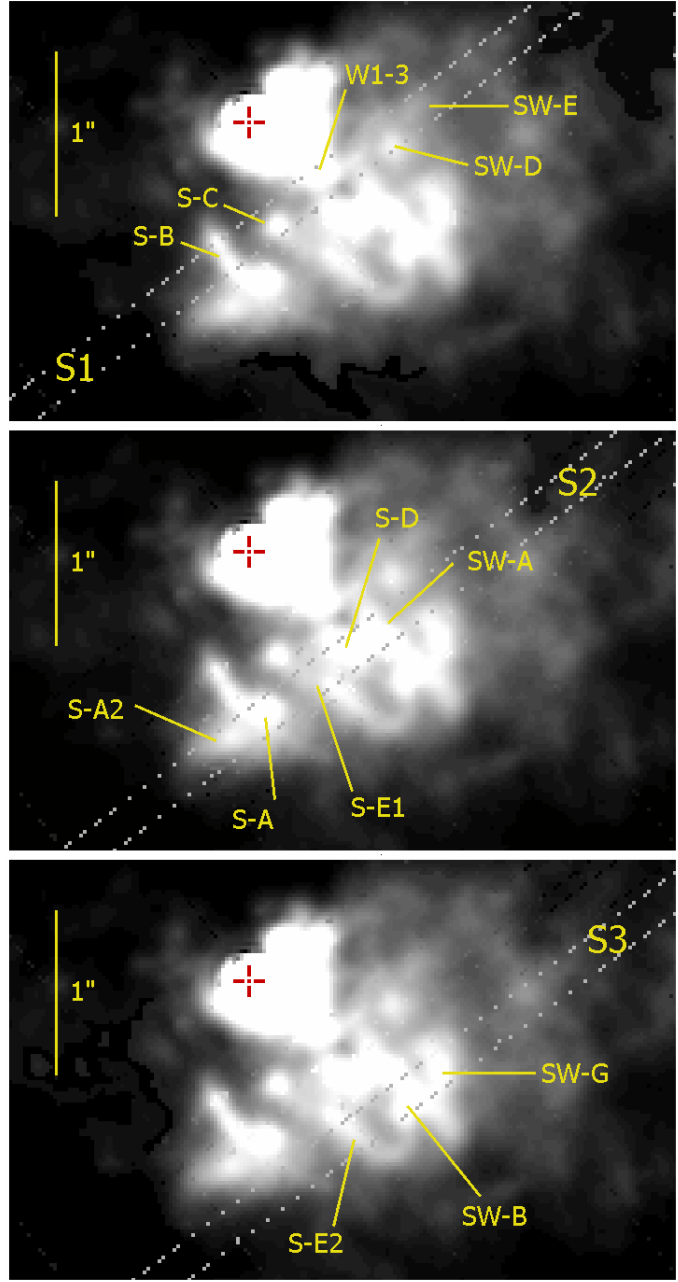}
\caption{Locations of the three ``S'' slits on the red F656N image. The 
separate knots discussed here and listed in Table 2 are identified. The central star is marked by a red cross.}
\end{figure}

\begin{figure}
\figurenum{5}
\plotone{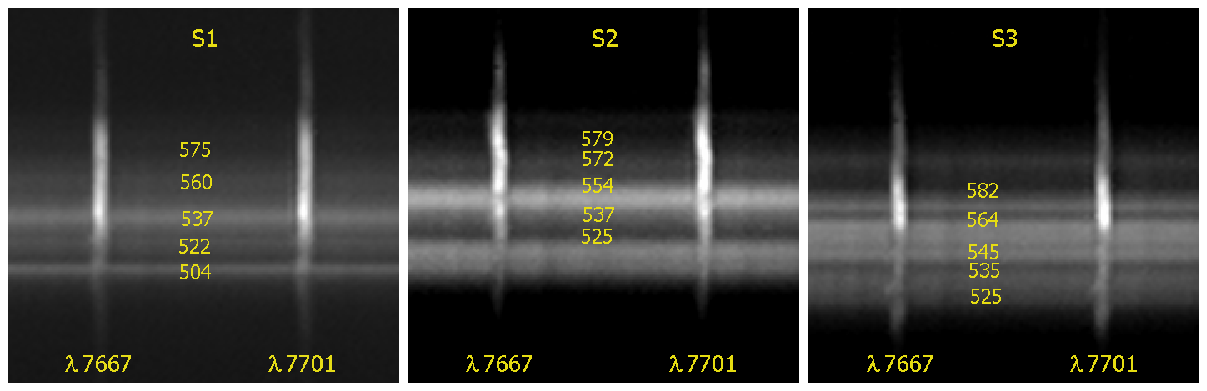}   
\caption{The two-dimensional images along the three slits showing the K I emission perpendicular to the dispersion with the row numbers in pixels corresponding to the profiles in Figure 6.}
\end{figure}

The profiles in Figure 6 illustrate the shift in the K I Doppler velocities of the knots along the slits relative to the K I velocity of the star shown as a dashed line. Unlike the complex profiles in \citet{RMH2005}, these profiles with the narrow slit and extracted on specific knots, are mostly symmetric and with a single velocity peak. The 2005 study used  a much wider 1$\farcs$0 slit which intercepted multiple features especially crossing the arcs in the outer ejecta. The only profile that shows evidence of multiple peaks is S3 row 525, but there is no obvious identification with a condensation. Close inspection of S3 in Figure 5 shows a blueward extension on the K I emission at that row. The relatively  bright extended nebulous feature, W1 knot C  contributes to both slits W1 and S1. The profile in Figure 6 is from S1, but  spectra were extracted from both slits and their measurements are included in Table 2. 

\begin{figure} 
\figurenum{6} 
\plotone{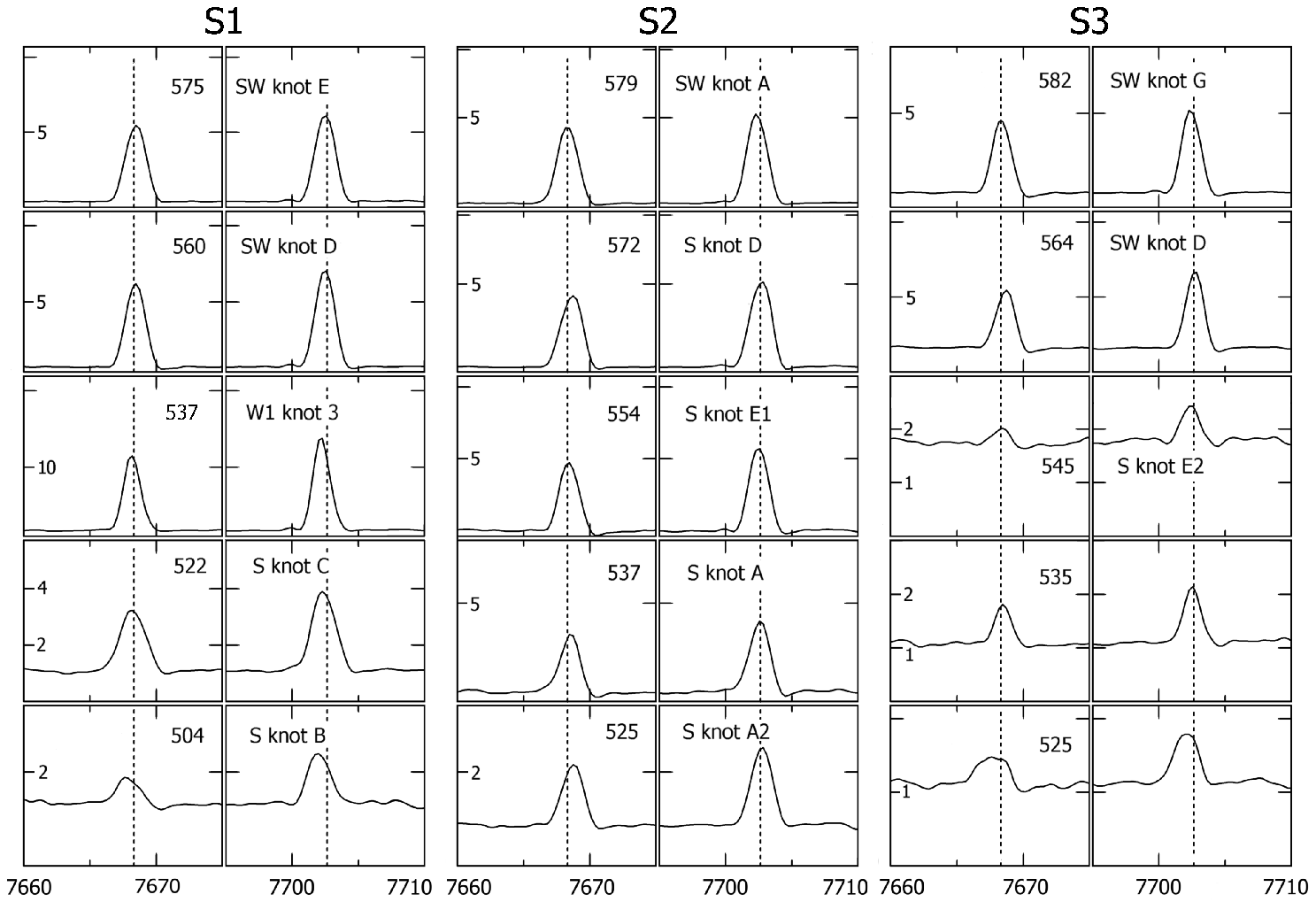} 
	\caption{The K I emission line profiles extracted along the slits. The row number (Figure 5) is given in the panel for the $\lambda$ 7667 line and the knot identification  in the neighboring profile for the $\lambda$ 7701 line. The flux calibrated spectra are in units of 10$^{-14}$ Watts cm$^{-2}$ vs the vacuum wavelength in {\AA}. }
\end{figure}

The measured Doppler velocities of the two K I emission lines are given in Table 2 with the velocity of the K I absorption feature if present, and the average velocity of the absorption lines, ith the number of lines in parenthesis, from  the  spectrum with 
grating tilt 8561. The radial velocity is the average velocity of the 
K I lines relative to the velocity of the K I lines in the star's spectrum 
measured in the 0$\farcs$1 slit (Table 1).

We determine the  transverse velocity from the proper motions of the knots measured in  
the two epochs of HST/WFPC2 images from 1999 and 2005 as decribed in \citep{RMH2007}.  VY CMa was imaged in four filters, F410M, F547M, F656N, and F1042M, 
with a range of exposure times.   The images are tightly aligned in pixel space and  blinked to identify any offset. We measure the x and y positions of the blinked images three times in each filter combination in which the knot or filament is identified. The measurements  are then combined for a weighted mean. 

For this paper we checked all of our previous measurements, and in some cases repeated them to resolve discepancies or multiple entries from the different 
filters. These are also noted in the footnotes. In all cases, the previous
measurements were adjusted to the distance change from 1.5 kpc used in \citep{RMH2007} to 1.2 kpc \citep{Zhang} and the change in the adopted velocity for the central star, 55 km s$^{-1}$, see Paper 1.

Combining the resulting transverse velocity with the radial velocity from the K I lines relative to the star, we  then determine each 
knot's   orientation ($\theta$) with respect to the plane of the sky, the 
direction of motion ($\phi$) and total space velocity  and  then estimate its age or time since ejection assuming a constant velocity.  Although the winds of evolved cool stars are known to accelerate, these knots and arcs are discrete ejections, and it is not known if  the plasma is accelerated or slowed by interaction with  surrounding material.

 (Acceleration and/or deceleration of the knots obviously affect these 
  constant-velocity estimates.   With credible parameters, however, the 
  estimated times since ejection are altered by only a few percent.  
  For instance, consider two simple models with an escape speed between 
  55 and 90 km s$^{-1}$ just above the stellar surface, at $r \approx 8$ AU \footnote{The escape velocity depends on the mass of the star. Based on its luminosity, VY CMa had an initial mass of $\sim$ 30 M$_{\odot}$, possibly as high as 35-- 40. It has had a very high mass loss rate as a RSG and substantial mass loss on the Main Sequence. It therefore may have shed half its mass. Assuming a current 20 M$_{\odot}$, gives 70 km s$^{-1}$ for the escape speed.}. 
  In one model, suppose that the ejecta were accelerated instantaneously 
  in that region, and later were gravitationally decelerated as they moved 
  outward.  At radial distances of 300--800 AU, the true age since ejection 
  is then about 10\% smaller than the time estimated by our simplified  
  constant-velocity estimate.  Alternatively, suppose that the ejection 
  process was quite gradual, with a net outward acceleration proportional  
  to $1/r^2$.  In that case, the true age is about 5\% longer than the  
  constant-velocity estimate.  Further details are beyond the scope of 
  this paper, but most likely ejection scenarios would produce results 
  intermediate between these two cases.)

We note that most of these  knots or features relatively near the star have approximately the same total velocity, $\sim$ 20 - 30 km s$^{-1}$, and to a first order their age or time since ejection is correlated with distance from the star
which is not surprising.  Based on their orientations and directions of motion the condensations to the southeast along S1 and  S2 (S 
knots B, C, A and A2) initially look as though they may be physically associated and from the same ejection event. But they were apparently ejected at different times with  ages that correlate with three active periods  about 120, 200, and $\approx$  250 yrs ago also indicated for several of the other knots and condensations.  

In contrast, the  knots to the south and west along S2 and S3  appear to be 
clustered around what looks like a darker patch or more correctly a region 
with low emission. These condensations are identified with an infrared bright feature called the SW Clump \citep{Shenoy,Gordon} and the knots were referred to as the SW knots in \citet{RMH2007}.

\begin{deluxetable*}{llllllllllll}
\tabletypesize{\footnotesize}
\rotate
\tablenum{2}
\tablecaption{Summary of Measured Heliocentric Velocities, Motions and Ages}
\tablewidth{0pt}
\label{tab:Meas}
\tablehead{
\colhead{Object} & 
\colhead{K I em Vel} & 
\colhead{K I abs Vel} & 
\colhead{Abs. Lines Vel.} & 
\colhead{Radial Vel.\tablenotemark{a}} & 
\colhead{Trans Vel. } & 
\colhead{Total Vel.} &
\colhead{Direct. $\phi$} &
\colhead{Orient. $\theta$} & 
\colhead{Dist.} & 
\colhead{Age} &  
\colhead{Comment} \\
\colhead{Name} &
\colhead{km s$^{-1}$}  &
\colhead{km s$^{-1}$}  &
\colhead{km s$^{-1}$}  & 
\colhead{km s$^{-1}$}  &
\colhead{km s$^{-1}$}  &
\colhead{km s$^{-1}$}  &
\colhead{deg} &
\colhead{deg} &
\colhead{AU} &  
\colhead{yr} &
\colhead{} 
} 
\startdata 
S knot B &  25.4, 30.0 & \nodata, -43.6 & 75.0 $\pm$ 9.9(16) & -26.8 $\pm$ 1.4 & 30.2 $\pm$ 2.5 & 40.4 $\pm$ 2.8 & 148 $\pm$ 7 & -41.6 $\pm$ 3 & 1020 & 122 &  slit S1 \\
S knot C &  41.5, 43.6 & \nodata, \nodata & 68.8 $\pm$ 9.4 (15) & -12.2 $\pm$ 2.2 & 18.2 $\pm$ 1.0 & 21.8 $\pm$ 2.4 & 151 $\pm$ 12 & -33.8 $\pm$ 8 & 876 & 194 & slit S1 \\
W1  knot C \tablenotemark{b} & 42.2, 41.6 & \nodata, -22.5 & 79.9 $\pm$ 9.1(17) & -13.5 $\pm$ 3.2 & 21.9 $\pm$ 2.7 & 25.7 $\pm$ 4.2 & -134 $\pm$ 4 & -31.6 $\pm$ 9 & 636 & 120 &  slit S1 \\ 
   \nodata                 & 46.1, 44.8 & -12.1, -18.3 & 72.9 $\pm$ 8.4 (15) & -8.6 $\pm$ 4.0 & \nodata & 23.5 $\pm$ 5.0 & \nodata & -21.4 $\pm$ 10 & \nodata & 130 & slit W1 \\ 
SW knot D &  55.2, 53.0  &  \nodata, -21.4 & 95.9 $\pm$ 11.0 (15) & -1 $\pm$ 3.8 & 23.5 $\pm$ 3 & 23.5 $\pm$ 4.8 & -63 $\pm$ 11 & -2.4 $\pm$ 10 & 993 & 205 & slit S1 \\
SW knot E &  55.2, 53.7 & \nodata, \nodata  & 86.1 $\pm$ 13.9 (13) &   -0.5 $\pm$ 2 &15.6 $\pm$ 4.5 & 15.6 $\pm$ 5 & -116 $\pm$ 10 & -1.8 $\pm$ 10 & 1260  & 390  & slit S1 \\ 
S knot A2 \tablenotemark{c} & 68, 64 & \nodata, \nodata & 90.5 $\pm$ 8.6 (16) & 11 $\pm$ 4.2 & 25.2 $\pm$ 5.5 & 27.5 $\pm$ 7 &  135 $\pm$ 12 & 21.8 $\pm$ 10  & 1380 & 243 & slit S2 \\
S knot A  &  58.7, 58.4 & \nodata, \nodata & 94.3 $\pm$ 8.0 (15) & 4.4 $\pm$ 2.6  & 29.4 $\pm$ 2.4  & 29.7 $\pm$ 5 & 177 $\pm$ 5 & 8.5 $\pm$ 8 & 1170 & 189 & slit S2 \\
S knot E1 \tablenotemark{d} & 52.8, 53.3 & \nodata, -20 & 96.9 $\pm$ 10.1 (15) & -2.0 $\pm$ 4.2 & 20.1 $\pm$ 7.2 &  20.2 $\pm$ 7.4 & 77 $\pm$ 16  & -5.6 $\pm$ 16 & 1156 & 276 & slit S2, SW Clump \\
S knot D  \tablenotemark{e} & 65.7, 64.6  & \nodata, -23.4 & 88.1 $\pm$ 13  (13) & 10.2 $\pm$ 5  &  17.1 $\pm$ 4.7 &  19.9 $\pm$ 6.9  &  -120 $\pm$ 8 &  27 $\pm$ 10 & 1016 & 247 & slit S2, SW Clump  \\
SW knot A  &  48.9, 47.1 & \nodata, -28 &  94.9 $\pm$ 8.5 (14)  & -7.0 $\pm$ 5 & 24.1 $\pm$ 2.5 & 25.1 $\pm$ 5.5 &  -101 $\pm$ 23 & -16.2 $\pm$ 10 & 1200 & 231 & slit S2, SW Clump \\
S knot E2 \tablenotemark{d}  & 54.0, 52.5 & \nodata, -18.7 & 84.5 $\pm$ 8.7 (16) & -1.75 $\pm$ 3.5 & 19.1 $\pm$ 3.5 & 19.2 $\pm$ 5 & 146 $\pm$ 5 & -5.2 $\pm$ 10 & 1380 & 348 & slit S3, SW Clump \\
SW knot B  \tablenotemark{f} & 64.2, 62.7 & \nodata, \nodata & 95.2 $\pm$ 11.0 (16) & 8.5 $\pm$5 & 21.5 $\pm$ 5 & 23 $\pm$ 7 & -47 $\pm$ 10 & 21.6 $\pm$ 10 & 1460 & 308 & slit S3, SW Clump \\
SW knot G \tablenotemark{g} & 50.0, 48.3 & \nodata, -22.2 & 93.1 $\pm$ 9.2 (16) & -5.9 $\pm$ 2.5 & 26.5 $\pm$ 3 & 27.1 $\pm$ 4 & -143 $\pm$ 4 &  -12.5 $\pm$ 6 & 1490 & 264 & slit S3 \\ 
\enddata
\tablenotetext{a}{The average K I em velocity relative  to the K I emission line velocities measured on the star with the 0$\farcs$1 slit. See Table 1 and Paper I.} 
\tablenotetext{b}{This condensation is measured in Slits S1 and W1.}
\tablenotetext{c}{New measurement, not included in \citet{RMH2007}.}
\tablenotetext{d}{S knots E1 and E2. This feature was measured but not included in \citet{RMH2007}. The faint extension, E1, on slit S1 and the brighter E2 in S3 were measured separately. The large uncertainties on E1 are due to its faintness.}
\tablenotetext{e}{Two features D1 and D2 were identified in \citet{RMH2007} with discrepant measurements in different filters. Remeasured.}
\tablenotetext{f}{Remeasured.}
\tablenotetext{g}{Discrepant measurments in \citet{RMH2007}. Remeasured.} 
\end{deluxetable*}

\subsection{The SW Clump}

The infrared bright feature visible in the longer wavelength images from 1 to 10 $\mu$m to the southwest of the star is referred to as the ``SW Clump'' in several papers \citep{RMH2007,Shenoy,Gordon}. Several knots and diffuse condensations
 identified with this feature in Table 2 appear to be spatially distributed 
around a region of lower flux that appears dark in the visual images. Figure 7 shows the HST 1$\mu$m image with the Ks (2.15$\mu$m) image from LBT/LMIRCam reproduced from \citet{Shenoy}. The visual knots are marked on the 1$\mu$m images and possible cross identifications are shown on the Ks image.  Note that the 2$\mu$m image shows the same basic structure as the 1$\mu$m  picture with knots distributed in a rather elongated shape  around a region of lower emission. Comparison of the superposed HST visual and 1$\mu$m images, suggests that most of the near-IR flux appears to be coming from S knot D and SW knot A, which are also the two largest knots and the closest to the star. Thus separate knots may contribute to the strong IR flux identified as the SW Clump.  

\begin{figure}[h] 
\figurenum{7} 
\epsscale{1.1} 
\plottwo{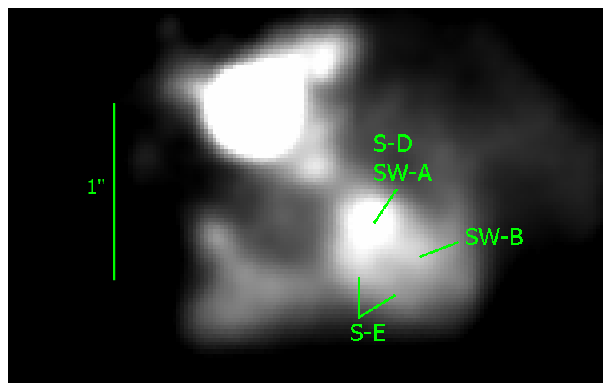}{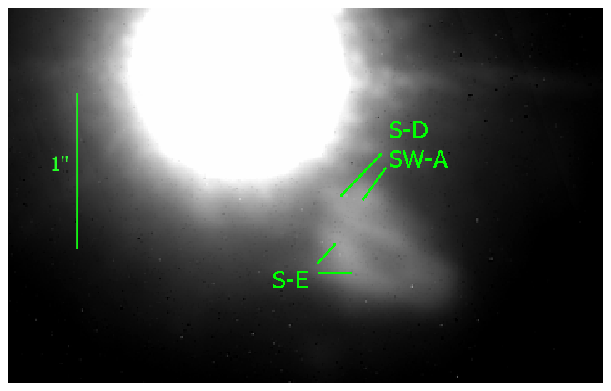}
\caption{Left: The HST F1042 image with the positions of the visual knots identified. Knots S-D and SW-A are merged in the 1$\mu$m image. Right: The Ks band images from \citet{Shenoy}. Possible cross identifications of the partially resolved features from the visual images are marked.}
\end{figure}

The four or five knots identfied in the visual image (Figure 4) appear to be distributed in an arc or possible elliptical shape around the ``dark'' lower density region. This is suggestive of an expanding large knot or bubble with the knots  on the outer rim. The apparent non-radial motion (with respect to the star) specifically of knots S-E2 and SW-B support a possible expanding bubble. This reminds us  of the nearly-circular arcs in the inner ejecta of the warm hypergiant IRC~+10420 
that were shown to be expanding bubbles \citep{Tiffany}. 

To test this possibility, we fit an ellipse to the positions of the knots in 
the two  epochs, 6.23 years apart. 
  Comparing these loops or arcs via cross-correlations would be 
difficult because their inner knots are much brighter than their   
outer parts. Faint knot S-E1 was excluded because its position is relatively 
uncertain.  Thus, in order to have five points necessary to define 
an ellipse, we used a point on the south rim, part of a luminous arc 
on the apparent south side of the "SW clump."  Its position is less 
well defined than the other knots and it has no velocity measurement. 
Obviously it would be better to use more than five spatial points, 
but no other features were suitable.  Since the intrinsic deviations 
from a true ellipse are presumably similar for both epochs, we can 
meaningfully compare these two ellipses. 

The best-fit ellipse is shown superposed on the F656N image from 1999 in Figure 8. The parameters from the ellipse fits are summarized in Table 3. The two ellipses coincide with a r.m.s. 
difference of 16 mas when we allow for expansion, and the 
disagreement in the radial direction is of the order of 5 mas; 
this is satisfactory considering the fuzziness of each data point.

\begin{figure}[h]
\figurenum{8}
\epsscale{0.7}
\plotone{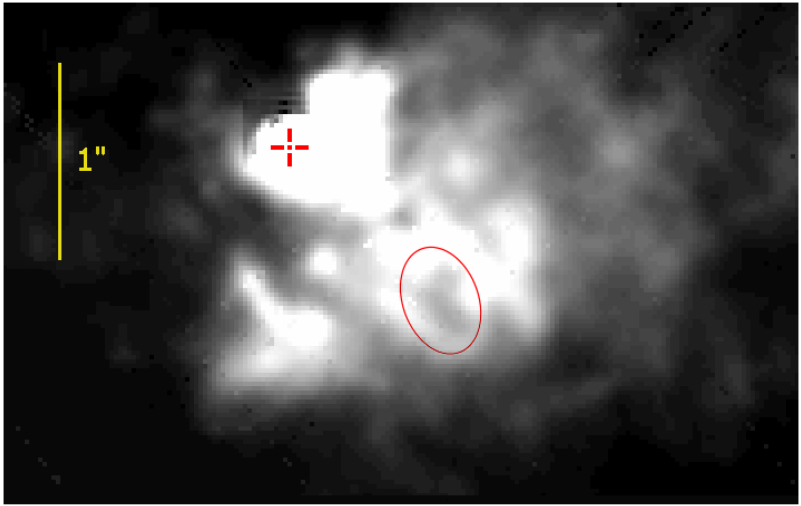}
\caption{Ellipse fit to the  knots possibly associated with the SW Clump shown superposed on the 1999 red F656N image.}
\end{figure}

\begin{deluxetable*}{lllll}
\tabletypesize{\footnotesize}
\tablenum{3}
\tablecaption{Ellipse Fits}
\tablewidth{0pt}
\label{tab:ellipse}
\tablehead{
\colhead{Epoch} & 
\colhead{Major axis (AU)} & 
	\colhead{Major Axis PA}  & 
	\colhead{Center Dist (AU)} &
	\colhead{Center PA} 
}
\startdata 
1999.22 & 0\farcs562 (675)   & +20 deg   & 1\farcs090 (1310) & +44 deg \\
2005.45 & 0\farcs575 (691) & +12 deg   &  1\farcs123 (1350) & +43 deg      
\enddata
\end{deluxetable*}

The centers of the two ellipses are  consistent with an 
expansion age of about 200 years relative to 1999 (projected velocity 
$\sim$ 25 km s$^{-1}$), while the major-axis expansion similarly 
indicates about 260 years.  These values are comparable  
with other features south and southwest of the star. 
Although the visual images and the ellipse fits are encouraging there are problems with this model for 
S knot D and SW knot A. Both have a direction of motion consistent with 
radial motion outward from the star as opposed to an expanding outer shell of a bubble moving away from the ellipse center. The positive radial velocity for S-D shows it is projected backward, away from the plane of the sky, while SW-A is 
forward, in front of the plane of the sky.  

An alternative possibility would place knots S-E2 and SW-B and the south rim on
an expanding arc or loop  like several small arcs
seen throughout the ejecta as well as the giant arcs/loops in the outer ejecta.
SW-B is projected away from us but its orientation is consistent with the far side of a loop tilted with respect to our line of sight. If that is the case, this arc or loop would have been ejected about 300 years ago, and the S-D and SW-A knots would then be chance superpositions from later eruptions. 

Thus it is not clear if the SW clump is a coherent object, or chance alignment 
of unrelated clumps of dust and gas. The 2-5$\mu$m imaging by \citet{Shenoy} 
still puts strong constraints on the feature. The high surface brightness 
 requires that the illuminated clump(s) have a relatively clear line of sight 
 to the star and be optically thick to scattering by dust. The imaging polarimetry \citep{Shenoy15} constrains the albedo to 40\% or less, otherwise multiple scattering would reduce the observed fractional polarization at 3.1$\mu$m below the observed value. These observations combined with the optically thick emission observed at 
 11$\mu$m \citep{Gordon} require that a minimum mass of 5 $\times 10^{-3}$ M$_{\odot}$  must be present in the knots, whether they are kinematically related or not.

Another example is shown in a  recent paper on high resolution imaging with ALMA \citep{Asaki}. Images  at $\sim$ 10 AU (8-mas) resolution show that ALMA Clump 
C in VY CMa is resolved into many small condensations of higher and lower intensity making up the total flux density detected at lower resolution.  The 
condensations have  a range of apparent sizes, but  most  appear to be much 
smaller than the knots in the visual and and near-IR images discussed here, 
although Clump C is much 
younger than the SW Clump (see \S {5}), and may expand with time.

\section{The Mass Loss History}

The most important conclusions from our previous studies of VY CMa \citep{RMH2005,RMH2007,TJJ} were not only the very visible evidence for high mass loss episodes, but the massive outflows in different directions ejected at different times over several hundred years, suggestive of significant periods of surface activity. With the improved spatial 
resolution from the STIS spectra, we've determined the ejection ages for the 
separate knots within about one arcsec of the star. The closest, along  slits W1, W2 and E, were apparently ejected within the past century (Paper I), and the inner knot described in this paper was ejected $\approx$ 1985 -- 1990. The numerous 
knots to the S and SW including those associated with the SW Clump, have ejection ages which concentrate around 120, 200, and 250 years ago. 

For comparison with the large arcs and features visible in the more distant ejecta \citep{RMH2007}, we've redetermined their ejection ages (Table 3) with the improved distance to VY  CMa \citep{Zhang} and our radial velocity for the central star from Paper I and see Table 1.  The revised ages 
average slightly less than reported in our 2007 paper.  Our mid and far-infrared observation with SOFIA \citep{Shenoy} found no extended cold dust with an age greater than about 1200 years. Combined with our 
current measurements, VY CMa thus has a history of discrete mass loss events over the past 1000 years which appears to be continuing.

\begin{deluxetable*}{lc}
\tabletypesize{\footnotesize}
\tablenum{4}
\tablecaption{Revised Ejection Ages for Major Features}
\tablewidth{0pt}
\label{tab:Ages}
\tablehead{
\colhead{Feature} & 
\colhead{Ejection Age (yr)}  
} 
\startdata 
Northwest Arc & 534   \\
Arc 1         & 637   \\
Arc 2         & 400   \\
West Arc      & 312   \\
South Arc     & 373   \\
SE Loop       & 271   \\ 
Outer ``spikes'' &  1000--1500 
\enddata
\end{deluxetable*}

In Figure 9, we show VY CMa's light curve from 1800 to 2000. The upper panel for the 19th century is from the table in \citet{Robinson71}. We use the magnitudes from his column 4 which are reduced to the Harvard photometry and are approximately visual magnitudes. The second panel is from \citet{Robinson70} and are blue magnitudes measured from the Harvard patrol plates\footnote{VY CMa is not yet in the on-line database for the DASCH digitization of the Harvard plate collection.} from about 1900 to 1950. The points from $\sim$ 1963 -- 1970 are from the Remeis Observatory,  see \citet{Robinson70}. The bottom panel shows the more recent visual magnitudes from the AAVSO. 

Several features are notable. VY CMa was nearly naked-eye
at about 6 - 6.5 magnitudes at the start of the 19th century until a major period of variability from $\approx$ 1870 -- 1880 with multiple dimmings.  There is a 20 year gap from $\sim$ 1850 - 1870, but by 1872 the star is a magnitude fainter. There are three recognizable minima. The  decline to about 8th magnitude in 1880 is followed by a gap of 10 years. We do not know if the star recovered or had additional minima. But by 1890 - 1893, it was a magnitude below its unobscured  brightness. Fortunately the 20th century light curve, (middle panel) overlaps the 1890 - 1910 period with no additional deep minima and confirms the decline by 1 to 1.5 magnitudes to 7.5 -- 8 magnitudes in the visual (9.5 - 10 blue mag).  It has stayed near this fainter magnitude since.  Today VY CMa is typically 8 mag in the visual with a B-V color of $\approx$ 2.0. Since the the 19th century the light curves show two periods with significant variability, from $\approx$ 1925 to 1945 with dimming by two or more magnitudes in the blue and more recently from about 1985 to 1995 dimming by 1 to 1.5 magnitudes in the visual.

\begin{figure}[h] 
\figurenum{9} 
\plotone{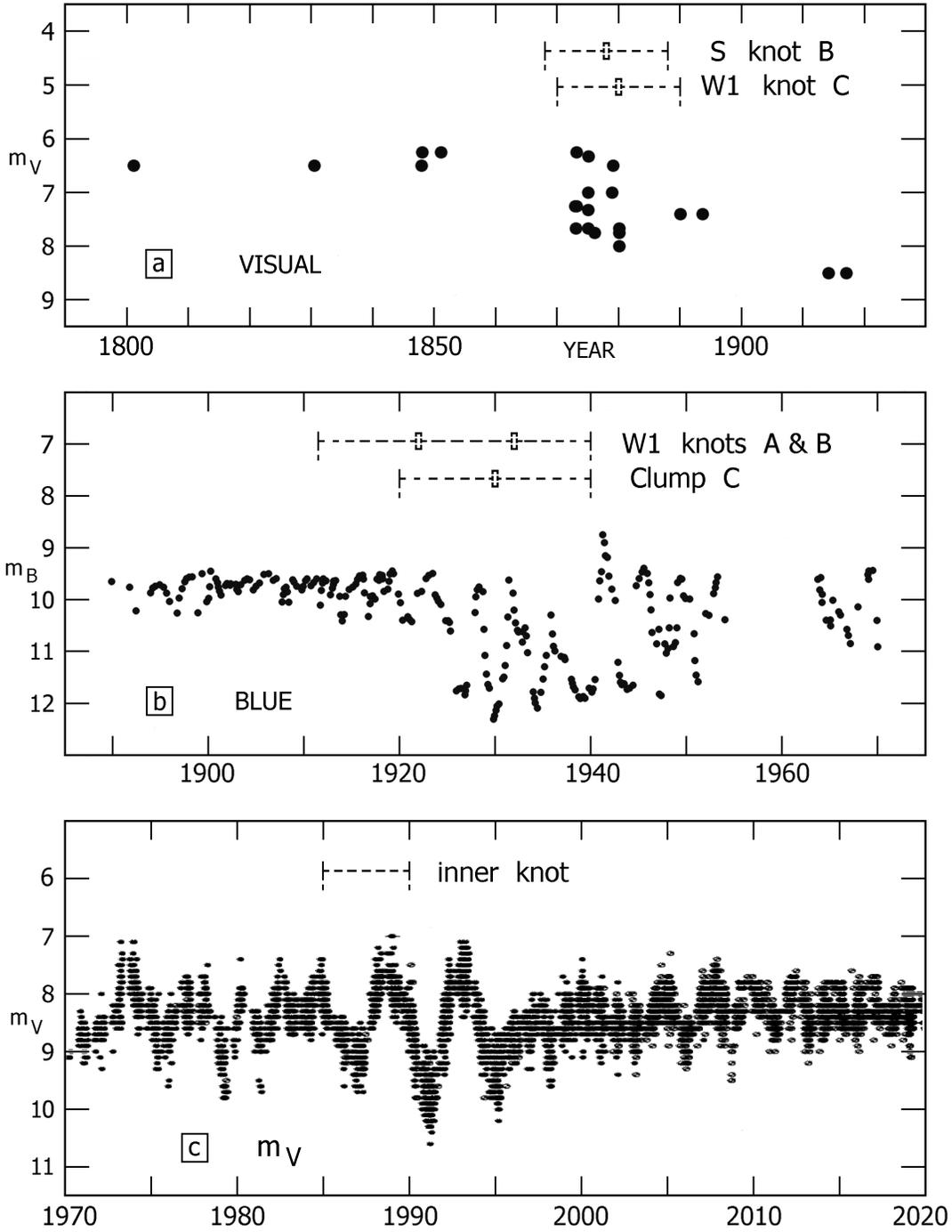}
\epsscale{0.9}
	\caption{The light curves with the probable ejection times times for the different knots. Upper panel (a):  the light curve for the 19th century \citep{Robinson71}. Middle panel (b): mid 20th century \citep{Robinson70}. Bottom panel (c); the most recent photometry from the AAVSO. We use the ages based an assumed constant velocity. Uncertainty due to acceleration or deceleration is small \S {4}, comparable to the error bars from other sources.}
\end{figure}

The ejection ages of several of the knots are marked  on the light curves. We emphasize that they agree with these extended periods of variability and change. The most recent, the inner knot visible in the narrow slit on the star is consistent with an ejection during VY CMa's most recent active period. The knots discussed in Paper I, along slit W1 have ejection dates from $\approx$ 1920 to 1940 corresponding to the long very active period.

In a recent paper on the dusty obscured clumps discovered with ALMA,  \citet{Kaminski} discussed the masses, ages and other properties of the condensations in this prominent feature just to the east of the star. Kaminski also compared his estimated ages with the 20th century light curves shown here in  Figure 9. Unlike our results, he concluded that there was no correlation between their ages and
the extended periods of variability. However, he adopted a rather high expansion velocity of 60 km s$^{-1}$ for the clumps. Velocities this high are only measured for a couple of the most distant arcs \citep{RMH2007}. The average outflow velocity for the small knots close to the star (Paper I) and at comparable projected distances as the ALMA clumps, is 27.5 km s$^{-1}$.  With this adopted velocity we estimate an age of about 70 years for Clump C, for example, corresponding
with the extended period of large light variations in the first half of the 20th century, Figure 9.

The most interesting  period may be  1870 -- 1880  and the post-1880 fading from which VY CMa has not recovered. S knot B and W1 knot C plus possibly the W2 knot have ejection ages that correspond to this period. Their different locations and orientations with respect to the star suggest that surface activity occurred over much of the star with separate outflows in different directions lasting  at least 10 years.  The  major dimming of the star  after 1880 by one or more  magnitudes  is very likely the origin of the present obscuration of the central star,
 the ``VY'' peak in the ALMA images. One of the unexpected conclusions in Paper I was the requirement that the visible compact knots just to the west of the star have a relatively clear line of sight to the star to explain their very strong K I and molecular emission. We thus suggested  that there must be a hole or gaps in the surrounding circumsteller ejecta near the star. Alternatively, the very dusty `VY' may not completely surround the star and is not obscuring it in all directions, but is instead a large knot, possibly similar to Clump C in the ALMA maps, ejected $\sim$ 1880 nearly in our line of sight. This possibility also helps explain the strong K I lines observed in some of the knots to the south of the star suggesting a clearer line of sight to the star in other directions as well.

\subsection{Comments on Betelgeuse}

The famous naked-eye red supergiant Betelgeuse ($\alpha$ Ori) recently experienced an unexpected  visual dimming beginning in 2019 December and continuing well into 2020 \citep{Guinan19,Guinan20} reaching an unprecedented minimum in 2020 February, fading by about one magnitude. A visual image from 2019 December observed with SPHERE on the VLT showed a remarkable corresponding fading of its southern hemisphere \citep{Montarges}. At about the same time UV spectra from {\it HST/STIS} revealed remarkable changes with an increase in the UV flux and variations in the Mg II line supporting a corresponding outflow of material from the star \citep{Dupree}. The dimming in the light curve is attributed to dust, although other authors have questioned this \citep{Dharma,Harper}, and suggest that a change in the photospheric luminosity or a cooling of a large fraction of the photosphere due to dynamical effects may be responsible. 

The similarity of the correspondence between the periods of variability and dimming in VY CMa and in Betelgeuse with outflows  of gas is clear. Furthermore, high spatial near and mid- infrared imaging of Betelgeuse \citep{Kervella09,Kervella} reveals several clumps or knots of dusty material within one arcsec of the star. This current ejection  by Betelgeuse may be similar to the recent inner knot ejection observed in VY CMa's spectrum or to its 1920 -- 1940 active period, depending on how long its current variability lasts.  

With its very extended and complex ejecta, the much more luminous and more massive  VY CMa is clearly a more extreme example of high mass loss events extending over several hundred years. There is definitely a difference in scale and very likely in the energies involved. For example, the deep minima in VY CMa typically lasted about  3 years and longer in a couple of  cases, while the dimming in Betelgeuse only lasted about 150 days. Nevertheless,  the presence of outflows from the surface of the  more typical red supergiant Betelgeuse suggests that these discrete mass loss episodes are  not unique to the extreme hypergiants like VY CMa and IRC~+10420 \citep{Tiffany}. 
The evidence for similar activity in other red supergiants should be more fully
explored. The strong maser sources are obvious candidates \citep{Schuster,RMH2020}, and clumpy winds have been reported for $\alpha$ Sco \citep{Ohnaka} and $\mu$ Cep \citep{Montarges2}. Surface  activity and the accompanying  outflows may thus be a major contributor to mass loss in these red supergiants.  

\subsection{Outflows and Ejections:  Pulsation and/or Surface Activity?}

Betelgeuse is a known semi-regular variable with a period of $\approx$ 400$^{d}$ attributed to radial pulsations plus a longer $\approx$ 2000$^{d}$ period associated with convective/magnetic activity. Betelgeuse is also well known for large, bright regions associated with convective cells. \citet{Dupree} suggest that the outflow in Betelgeuse from a convective cell was enhanced by the outward motion in its 400$^{d}$ radial pulsation. The outflow expanded and cooled obscuring the S. hemisphere and possibly forming dust causing the deep dimming in its visual light curve.

VY CMa's light curve provides a somewhat different perspective with a complex history of extended periods of activity or dimming lasting 10 to 20 years separated by long periods of relative quiescence. VY CMa is not a known to show radial pulsatons.  It is an irregular variable.  

We find probable correlations of 2 to 3 separate mass ejections or outflows with the 19th century episode and with the 1920 -- 1940 period suggesting that each of the minima may be related to a separate outflow.    If the deep minima which last 3 years or more, are caused by dust formation, their longer duration suggest that the ejections are more massive, perhaps implying more dust and obscure the star for longer.  Except for post 1880 -- 1890, in each case, VY CMa returned to its previous brightness.  Our measurements show that the knots are moving in different directions with different spatial orientations, and thus will eventually move out of our line of sight. We have suggested here that the major and continued dimming after 1880, is due to  a dusty  outflow nearly in our line of sight.

Based on VY CMa's light curve, a recurring radial pulsation period seems an unlikely cause or enhancer for its multiple outflows and minima. In our previous papers on VY CMa and other evolved supergiants with high mass-loss episodes  such as IRC~10420,  we have attributed the separate outflows and ejections  to large scale surface and magnetic activity. By analogy with Betelgeuse, we expect that VY CMa will have extensive surface asymmetries, large convective cells or regions covering much of its surface, that could be responsible not only for major outflows but for the equivalent of coronal mass ejections but on a larger scale. Non-radial pulsation is a possibility, but how would one separate that from large
 active regions on a star with a radius of $\approx$ 7 AU?

\acknowledgments 
This work was supported by NASA through grant GO-15076 (P.I. R. Humphreys) from the Space Telescope Science Institute. L. Ziurys acknowledges support from NSF grant AST-1907910.  

\vspace{2mm}
\facilities{HST(STIS), AAVSO }


\end{document}